\documentclass[runningheads]{llncs}
\usepackage{graphicx}
\usepackage{amsmath}
\usepackage{color}
\usepackage{array}
\usepackage{multirow}
\usepackage[misc]{ifsym}
\begin{document}
%
\title{A Self-ensembling Framework for Semi-supervised Knee Cartilage Defects Assessment with Dual-Consistency}
\titlerunning{Semi-supervised Knee Cartilage Defects Assessment}

%

\author{Jiayu Huo\inst{1,2} 
\and Liping Si\inst{3} 
\and Xi Ouyang\inst{1,2}
\and Kai Xuan\inst{1}
\and Weiwu Yao\inst{3}
\and Zhong Xue\inst{2}
\and Qian Wang\inst{1}
\and Dinggang Shen\inst{2}
\and Lichi Zhang\inst{1}\textsuperscript{(\Letter)}
}
\authorrunning{J. Huo et al.}


%

\institute{Institute for Medical Imaging Technology, School of Biomedical Engineering, Shanghai Jiao Tong University, Shanghai, China \\
\email{lichizhang@sjtu.edu.cn} \\
\and 
Shanghai United Imaging Intelligence Co., Ltd., Shanghai, China \\
\and
Department of Imaging, Tongren Hospital, Shanghai Jiao Tong University School of Medicine, Shanghai, China\\
}

\maketitle              
\begin{abstract}
Knee osteoarthritis (OA) is one of the most common musculoskeletal disorders and requires early-stage diagnosis. Nowadays, the deep convolutional neural networks have achieved greatly in the computer-aided diagnosis field. However, the construction of the deep learning models usually requires great amounts of annotated data, which is generally high-cost. In this paper, we propose a novel approach for knee cartilage defects assessment, including severity classification and lesion localization. This can be treated as a subtask of knee OA diagnosis. Particularly, we design a self-ensembling framework, which is composed of a student network and a teacher network with the same structure. The student network learns from both labeled data and unlabeled data and the teacher network averages the student model weights through the training course. A novel attention loss function is developed to obtain accurate attention masks. With dual-consistency checking of the attention in the lesion classification and localization, the two networks can gradually optimize the attention distribution and improve the performance of each other, whereas the training relies on partially labeled data only and follows the semi-supervised manner. Experiments show that the proposed method can significantly improve the self-ensembling performance in both knee cartilage defects classification and localization, and also greatly reduce the needs of annotated data.

\end{abstract}
\section{Introduction}
Osteoarthritis (OA) is one of the most common joint diseases, which is characterized by a lack of articular cartilage integrity, as well as prevalent changes associated with the underlying bone and articular structures. OA can lead to joint necrosis or even disability if it is not intervened at an early stage \cite{karsdal2016disease}. Knee cartilage defects assessment is highly correlated to knee OA diagnosis \cite{peterfy2004whole}, so that it can be treated as a subtask. Magnetic resonance imaging (MRI) is a powerful tool for OA diagnosis. Compared with X-ray, MRI has a better imaging quality for cartilage and edema areas, which makes it practical for the early-stage clinical diagnosis.

Computer-aided diagnosis (CAD) based on MRI have achieved greatly for diagnosing OA, since it can reduce the subjective influences from the radiologists, and also greatly release the burdens of their works. A number of contributions have been achieved in the field of CAD using deep learning techniques \cite{antony2016quantifying,deng2009imagenet,liu2018deep}. For example, Antony \textit{et al.} \cite{antony2016quantifying} used a CNN model pretrained from ImageNet \cite{deng2009imagenet} dataset to automatically quantify the knee OA severity from CT scans. Liu \textit{et al.} \cite{liu2018deep} implemented a U-Net \cite{ronneberger2015u} for the knee cartilage segmentation, and fine-tuned the encoder to evaluate structural abnormalities within the segmented cartilage tissue. However, the good performance achieved by the supervised deep neural networks highly relies on the manually annotated data with extensive amount, which is generally high-cost. In order to alleviate the needs of huge amount manual annotations, several semi-supervised methods were developed. Laine \textit{et al.} \cite{laine2016temporal} designed a temporal ensembling model for the natural image classification. Yu \textit{et al.} \cite{yu2019uncertainty} proposed an uncertainty-aware framework for the left atrium segmentation. But, the semi-supervised framework for knee joint disease diagnosis has not been proposed yet. 

In this paper, we propose a self-ensembling semi-supervised learning approach, named as dual-consistency mean teacher framework (DC-MT), to resolve the high demand of annotated data. Our DC-MT framework aims to quantify the severity of knee cartilage defects simultaneously, to provide informative attention masks for lesion localization. This quantification task can be treated as a subtask of knee OA diagnosis. The attention masks highlight regions that related to knee cartilage defects and its severity can be used as the basis to interpret the diagnosis results in clinical practice. On the other hand, such attention-based localization tasks could improve the performance of knee cartilage defects classification.

In summary, the main contributions are listed as follows: 1) DC-MT consists of a student model and a teacher model, which share the same architecture. Two additional attention mining branches are added into the two models respectively to obtain the attention masks, which can be considered as the basis for classification. 2) We define an attention loss function to constrain the attention mask generation, which can yield more accurate attention masks. It could also let the classification results more credible if the corresponding attention masks are precise. 3) We propose novel dual-consistency loss functions to penalize the inconsistency of output classification probability and attention mask. It can help the whole framework achieve consistency between the student and teacher models in both attention and classification probability level, so that the two networks support each other to improve performance interactively.


\begin{figure}[!t]
\centering
\includegraphics[width=0.95\textwidth]{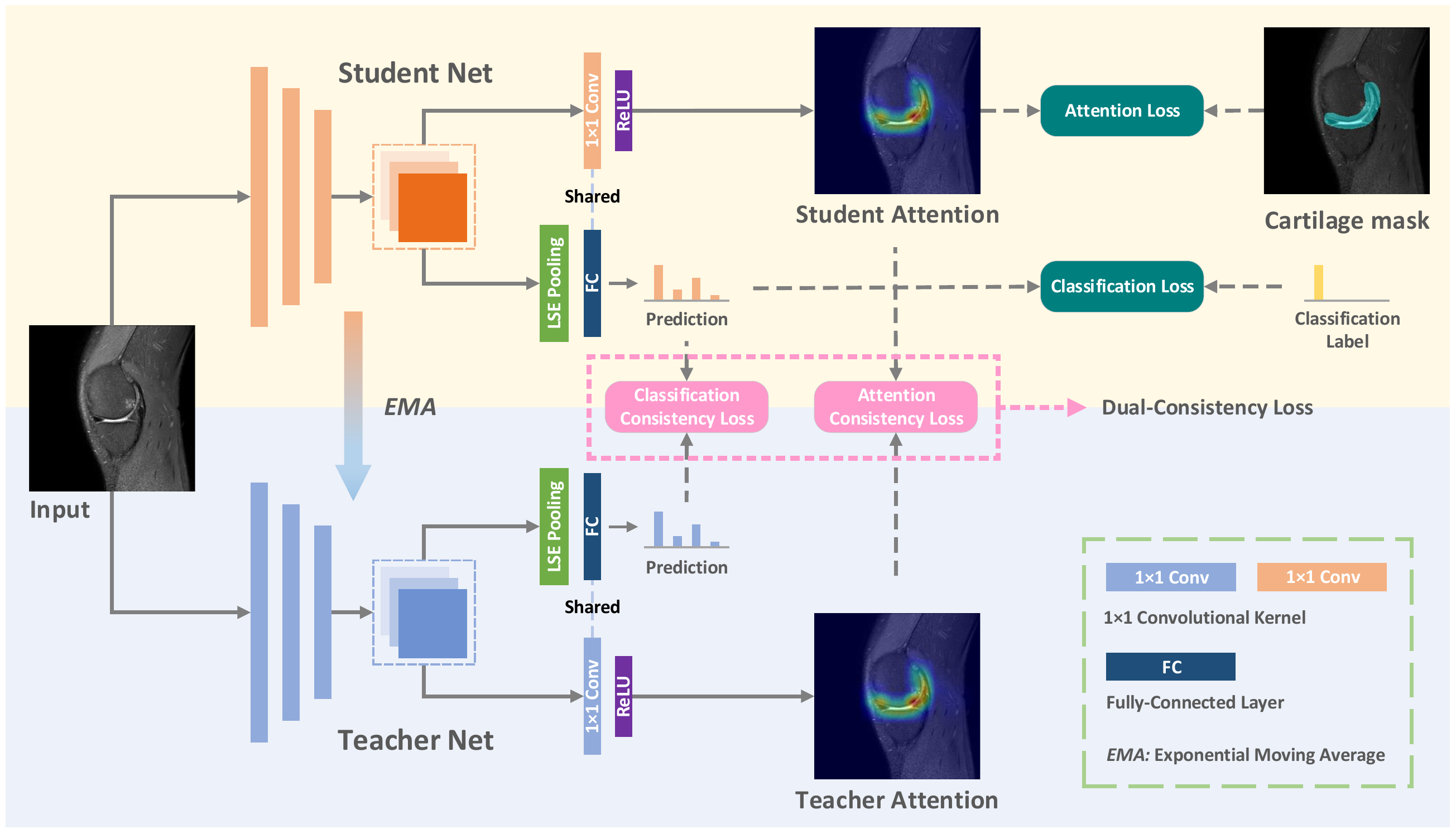}
\caption{The pipeline of our DC-MT framework for semi-supervised classification and localization of knee cartilage defects. Two dark green round rectangles denote the supervised loss functions, and two pink round rectangles denote the dual-consistency loss functions.} 
\label{fig:framework}
\end{figure}

\section{Methodology}
The proposed DC-MT framework for knee cartilage defects diagnosis is illustrated in Fig. \ref{fig:framework} , which consists of a teacher model and a student model with the same architecture. Both models generate the classification probabilities for knee cartilage defects severity and provide the attention masks for lesion localization simultaneously. The dual-consistency loss functions are proposed to ensure improved classification and localization performance.

\subsection{Mean Teacher Mechanism}
Mean teacher model \cite{tarvainen2017mean} is a self-ensembling model which is designed for the classification task of the natural image. It typically contains two models (i.e., student model and teacher model) with the same network structure. As shown in Fig. \ref{fig:framework}, a knee joint image is input to the student and teacher networks respectively. The output includes both the knee cartilage defects severity probabilities and the corresponding attention masks. Specifically, the student network is optimized by both the supervised and the unsupervised loss funtions, and the teacher model is updated by \textit{exponential moving average} (EMA) \cite{laine2016temporal}. The EMA updating strategy is used to merge network weights effectively through optimization. The weight of the teacher model $\theta_{\tau}^{'}$ at training step $\tau$ is updated by:
\begin{equation}
\label{eq:ema}
\theta_{\tau}^{'} = \alpha \theta_{\tau  - 1}^{'} + \left( {1 - \alpha } \right){\theta _\tau },
\end{equation}
where $\alpha$ is a decay factor that controls the weight decay speed, and $\theta_{\tau}$ is the student model's weight. It can be seen that the student network is more adaptive to training data and the teacher network is more stable. By using the two models, we hope that the final trained networks can demonstrate a combined advantage of the networks.

\subsection{Attention Mining}
The goal of attention mining is to generate attention masks while performing localization and classification tasks. In this work, the attention mining strategy is based on guided attention inference network \cite{li2018tell,ouyang2019weakly}. It shows that the generated attention masks will be more accurate if the segmentation results of the targets are added as the supervision. Here we apply a U-Net-based model to firstly segment the femur cartilage region and utilize it for attention supervision. Since the lesions are generally located in the cartilage region, it is indicated that our cartilage segmentation results can help refine the attention masks and improve their corresponding classification performance. In this way, we add an attention loss to constrain the attention mask generation. Besides, a regularization term is also added so that the attention mask which is small and within the segmented cartilage region is also acceptable. The entire attention loss is therefore defined as:
\begin{equation}
\label{eq:aloss}
{L_a} = {\lambda _a}\frac{{\sum\nolimits_k {{{\left| {{f_\theta }{{\left( {{x_i}} \right)}_k} - S{{\left( {{x_i}} \right)}_k}} \right|}^2}} }}{{\sum\nolimits_k {{f_\theta }{{\left( {{x_i}} \right)}_k} + \sum\nolimits_k {S{{\left( {{x_i}} \right)}_k}} } }} + {\lambda _r}\left( {1 - \frac{{\sum\nolimits_k {({f_\theta }{{({x_i})}_k} \cdot S{{\left( {{x_i}} \right)}_k})} }}{{\sum\nolimits_k {{f_\theta }{{({x_i})}_k}} }}} \right),
\end{equation}
where $f_{\theta}(x_{i})_{k}$ denotes the attention masks generated by the student model with input ${x_i}$ at the $k$-th pixel, and $S(x_{i})_{k}$ denotes the corresponding femur cartilage segmentation result. The U-Net-based model is denoted as $S$, and $\lambda_{a}$ and $\lambda_{r}$ are the loss weighting factors. With the help of the attention loss, the network can generate more accurate attention masks, which further improve the classification performance.

\subsection{Dual Consistency Loss}
Using the additional attention mining branch, the student model and teacher model yield a classification probability and an attention mask at the same time. To better coordinate the two networks, we need to ensure the consistency between output probabilities, and also between the attention masks. Hence, we propose the novel attention consistency loss to meet the requirement. When a batch of images are treated as input, the two models yield the probability and the attention mask, respectively. The student model is optimized by the supervision loss and the dual consistency loss, as a result the whole framework achieve a better performance. In this work, we design the dual-consistency loss functions as mean squared error (MSE) regards of probability and attention maps. Specifically, the dual-consistency loss functions are defined as:
\begin{equation}
\label{eq:ccloss}
\begin{aligned}
{L_{cc}} &= \frac{1}{n}\sum\nolimits_n {{{\left| {{p_\theta }({x_i})_n - {p_{{\theta ^{'}}}}({x_i})_n} \right|}^2}},
\end{aligned}
\end{equation}

\begin{equation}
\label{eq:acloss}
\begin{aligned}
{L_{ac}} &= \frac{{\sum\nolimits_k {{{\left| {{f_\theta }{{({x_i})}_k} - {f_{{\theta ^{'}}}}{{({x_i})}_k}} \right|}^2}} }}{{\sum\nolimits_k {{f_\theta }{{({x_i})}_k}}  + \sum\nolimits_k {{f_{{\theta ^{'}}}}{{({x_i})}_k}} }},
\end{aligned}
\end{equation}
where $\theta$ and $\theta^{'}$ represent parameters of the student and teacher models, respectively. $p_{\theta}(x_{i})$ and $p_{\theta}^{'}(x_{i})$ are probabilities of the models with respect to input $x_i$. $n$ represents the number of classification categories. With our proposed dual-consistency loss, the DC-MT framework can learn structure consistency and probabilistic distribution consistency synchronously, which is essential for the two models to support each other to improve the performance. 

The overall loss function consists of classification loss, attention loss and dual-consistency loss, which is shown as:
\begin{equation}
\label{eq:totalloss}
{L_{total}} = {L_c} + {L_a} + w_c(\tau ){L_{cc}} + w_a(\tau ){L_{ac}},
\end{equation}
where $L_c$ denotes the cross-entropy loss. $w_c(\tau )$ and $w_a(\tau )$ represent a ramp-up function of training step $\tau$ respectively, which can adjust the weighting factors of dual consistency loss functions dynamically. During the training procedure, the values of $w_c(\tau )$ and $w_a(\tau )$ will increase as the training procedure goes on. In our work, $w_c(\tau )$ and $w_a(\tau )$ are the same and set to $w(\tau )$. Here we define $w(\tau )$ as an exponential function, which is $w(\tau ) = {e^{ - 5 \cdot {{(1 - \tau /{\tau _{\max }})}^2}}}$). $\tau _{max}$ is the maximum training step. By this design setting, the network training procedure can be guided by the supervised loss at the beginning, so that the whole framework can be better trained, preventing the network sink into a degenerate condition.

\section{Experiments}
\subsection{Dataset}
In the experiments, we used 1408 knee T2 weighted MR images collected from Shanghai Jiao Tong University Affiliated Sixth People's Hospital to conduct the knee cartilage defects assessment task. The images were categorized into three classes according to whole-organ magnetic resonance imaging score (WORMS) \cite{peterfy2004whole}: normal thickness cartilage (WORMS 0 and 1), partial-thickness  defect cartilage (WORMS 2, 2.5, 3 and 4) and full-thickness defect cartilage (WORMS 5 and 6). An experienced radiologist selected and classified 6025 2D slices to generate the ground-truth, and the three categories are mostly balanced among them. Cartilage segmentation for all images was obtained through an inhouse U-Net toolkit, which was also validated by the radiologist. A dilation operation was applied to enlarge the segmentation results, which can reduce the difficulty of the localization task. We then randomly selected 90\% images of each class to form the training set, and the rest as the testing set. Particularly, the data selection was conducted according to subject, which can avoid slices from the same person were put into both the training and testing set.


\subsection{Experimental Settings}
The proposed algorithm was implemented using PyTorch. The backbone of the framework is the Se-ResNeXt50 model \cite{hu2018squeeze}. We changed the convolution stride in the fourth block so that a bigger feature map of the final convolution layer can be obtained. The size of the feature map is ${1 \mathord{\left/ {\vphantom {1 {16}}} \right. \kern-\nulldelimiterspace} {16}}$ of the input image size, which is necessary for accurate attention mask generation. Adam optimizer was employed and the value of weight decay was set to 0.0001. The learning rate was initialized with 0.001. The input image size of the network is ${\rm{256}} \times {\rm{256}}$, and data augmentation techniques were utilized to prevent over-fitting. The batch size was 30, including 20 labeled images and 10 unlabeled images. The loss weighting factors $\lambda_a$ and $\lambda_r$ in the attention loss were set to 0.5 and 0.001, respectively.

\subsection{Experimental Results}
\subsubsection{Efficacy of Attention Loss.}
We use four metrics to quantitatively evaluate the effect of the newly defined attention loss, including Recall, F1-Score, area under the ROC curve (AUC) and threshold intersection over union ratio (TIoU). TIoU means the ratio of the number of cases with correct localization against the total number of cases. If the intersection over union (IoU) ratio between the attention mask and the segmentation result is bigger than a prescribed threshold, the corresponding localization result is considered as correct. We set different thresholds T (T = \{0.1, 0.2, 0.3, 0.4, 0.5, 0.6, 0.7\}) and calculated IoU for evaluation. These values of IoU were then averaged to get TIoU. The first three metrics are used to evaluate the classification performance, and the last one for analyze the localization performance. We only use 10\% labeled training data to learn the student network.

A quantitative experiment of attention loss was conducted by setting the different values of  $\lambda_a$ and $\lambda_r$. The part of attention loss would not be calculated if the loss factor was set to 0. Table \ref{table:attention_loss_ablation} shows the result of the classification and localization performance under the different settings of the two attention loss factors. If $\lambda_a$ and $\lambda_r$ were both equal to 0, which means there is no supervision in attention mask generation, the network obtained a poor localization performance. However, if we only set the regularization item factor $\lambda_r$ to 0 and $\lambda_a$ to 0.5, the localization performance improved dramatically, also the classification performance was benefited and enhanced. With the help of the two penalties ($\lambda_a$ equals to 0.5 and $\lambda_r$ equals to 0.001), the network can achieve the highest performance in both classification and localization task. It also demonstrates the importance of attention loss when annotations are limited.

\begin{table*}[!t]
\caption{Attention loss ablation using the metrics of Recall, F1-Score, AUC and TIoU.}
\label{table:attention_loss_ablation}
\centering
\begin{tabular}{c|c|c|c}
\hline
Metrics & ${\lambda _a=0, \lambda _r=0}$ & ${\lambda _a=0.5, \lambda _r=0}$ & ${\lambda _a=0.5, \lambda _r=0.001}$ \\
\hline
Recall & 68.3\% & 74.4\% & \textbf{75.8\%} \\
F1-Score & 68.2\% & 74.7\% & \textbf{75.3\%} \\
AUC & 82.1\% & 86.0\% & \textbf{89.4\%} \\
TIoU & 7.5\% & 62.0\% & \textbf{71.3\%} \\
\hline
\end{tabular}
\end{table*}

\subsubsection{Evaluation of The Proposed Mechanism.}
This experiment illustrates the efficacy of our proposed mechanism. We trained the fully-supervised student network using all and 10\% labeled training data, which can be regarded as the upper-line and base-line performance, respectively. The proposed semi-supervised method also used all the training data, while certain percentage had their classification and segmentation information hidden. The experimental results are shown in Table \ref{table:exp_base}. It can be observed that the fully-supervised method achieved an average F1-Score of 75.3\% and TIoU of 71.3\% with only 10\% labeled data. By considering the feature consistency and structure consistency simultaneously and efficiently utilizing unlabeled data, our proposed mechanism further improved the performance by achieve 79.1\%, F1- score and 87.3\% TIoU. For the localization task, our method’s performance can reach the fully-supervised ones with all labeled data.

\begin{table*}[!t]
\caption{Comparison of Recall, F1-Score, AUC and TIoU between the fully supervised method and our proposed method. FS means full supervision and DC-MT is our proposed method.}
\label{table:exp_base}
\centering
\begin{tabular}{c|c|c|c}
\hline
Metrics & FS (10\% labels) & FS (100\% labels) & DC-MT (10\% labels)\\
\hline
Recall & 75.8\% & 85.0\% & 79.3\% \\
F1-Score & 75.3\% & 84.6\% & 79.1\% \\
AUC & 89.4\% & 93.7\% & 90.1\% \\
TIoU & 71.3\% & 90.3\% & 87.3\% \\
\hline
\end{tabular}
\end{table*}

We conducted another quantitative evaluation to analyze the importance of the attention consistency loss by adjusting the ratio of labeled data in the training set to obtain the labeled data contribution. The ratio of labeled data was set to 10\%, 30\% and 50\%, respectively. Moreover, we compared it with the original mean teacher model (MT) \cite{tarvainen2017mean} to prove the necessity of our proposed loss functions. Because the MT model was designed for semi-supervised classification tasks, we only compared the classification metrics for fair comparison. As shown in Table \ref{table:data_ablation}, an apparent improvement of the performance was observed as the ratio of labeled data increased. Here DC-MT (NAC) means that the attention consistency loss was not added into the proposed mechanism, and NAC stands for no attention consistency. Compared with the MT model, DC-MT (NAC) improved by 3.4\% Recall, 3.9\% F1-Score and 3.0\% AUC, respectively, when only 10\% labeled data were used for training. This demonstrates that the attention loss can help to improve the classification performance. When the attention consistency loss was added into the whole framework, DC-MT achieved 79.3\% Recall, 79.1\% F1-Score and 90.1\% AUC, which was the highest performance among all the methods. As the number of labeled data increases (e.g. from 30\% to 50\%), DC-MT (NAC) seemed to have reached a bottleneck. However, compared with DC-MT (NAC), DC-MT is still able to maintain stable growth in all these metrics. Although DC-MT achieved 91.9\% AUC when 30\% labeled data was used for training, which was lower than 92.7\% achieved by DC-MT (NAC), 83.1\% Recall and 83.2\% F1-Score of DC-MT were still higher than DC-MT (NAC). This also proved the importance of the novel attention consistency loss and the necessity of the combination between two attention related losses.

\begin{table*}[!t]
\caption{Quantitative analysis of all methods. DC-MT (NAC) means the attention consistency loss was not added into the proposed mechanism.}
\label{table:data_ablation}
\centering
\begin{tabular}{p{1.5cm}<{\centering}|p{2cm}<{\centering}|p{2cm}<{\centering}|p{2.3cm}<{\centering}|p{2.3cm}<{\centering}}
\hline
\multicolumn{2}{l|}{Metrics} & MT  & DC-MT (NAC) & DC-MT \\
\hline
\multirow{3}{*}{Recall} &10\% labels & 73.4\% & 76.8\% & \textbf{79.3\%} \\
&30\% labels & 78.0\% & 81.5\% & \textbf{83.1\%} \\
&50\% labels & 81.0\% & 81.3\% & \textbf{84.3\%} \\
\hline

\multirow{3}{*}{F1-Score} &10$\%$ labels & 72.7\% & 76.6\% & \textbf{79.1\%} \\
&30$\%$ labels & 78.0\% & 81.5\% & \textbf{83.2\%} \\
&50$\%$ labels & 81.0\% & 81.4\% & \textbf{83.8\%} \\
\hline

\multirow{3}{*}{AUC} &10$\%$ labels & 86.2\% & 89.2\% & \textbf{90.1\%} \\
&30$\%$ labels & 87.9\% & \textbf{92.7\%} & 91.9\% \\
&50$\%$ labels & 90.9\% & 92.7\% & \textbf{92.8\%} \\
\hline
\end{tabular}
\end{table*}

\subsubsection{Visualization Results}
\begin{figure}[!t]
\centering
\includegraphics[width=0.95\textwidth]{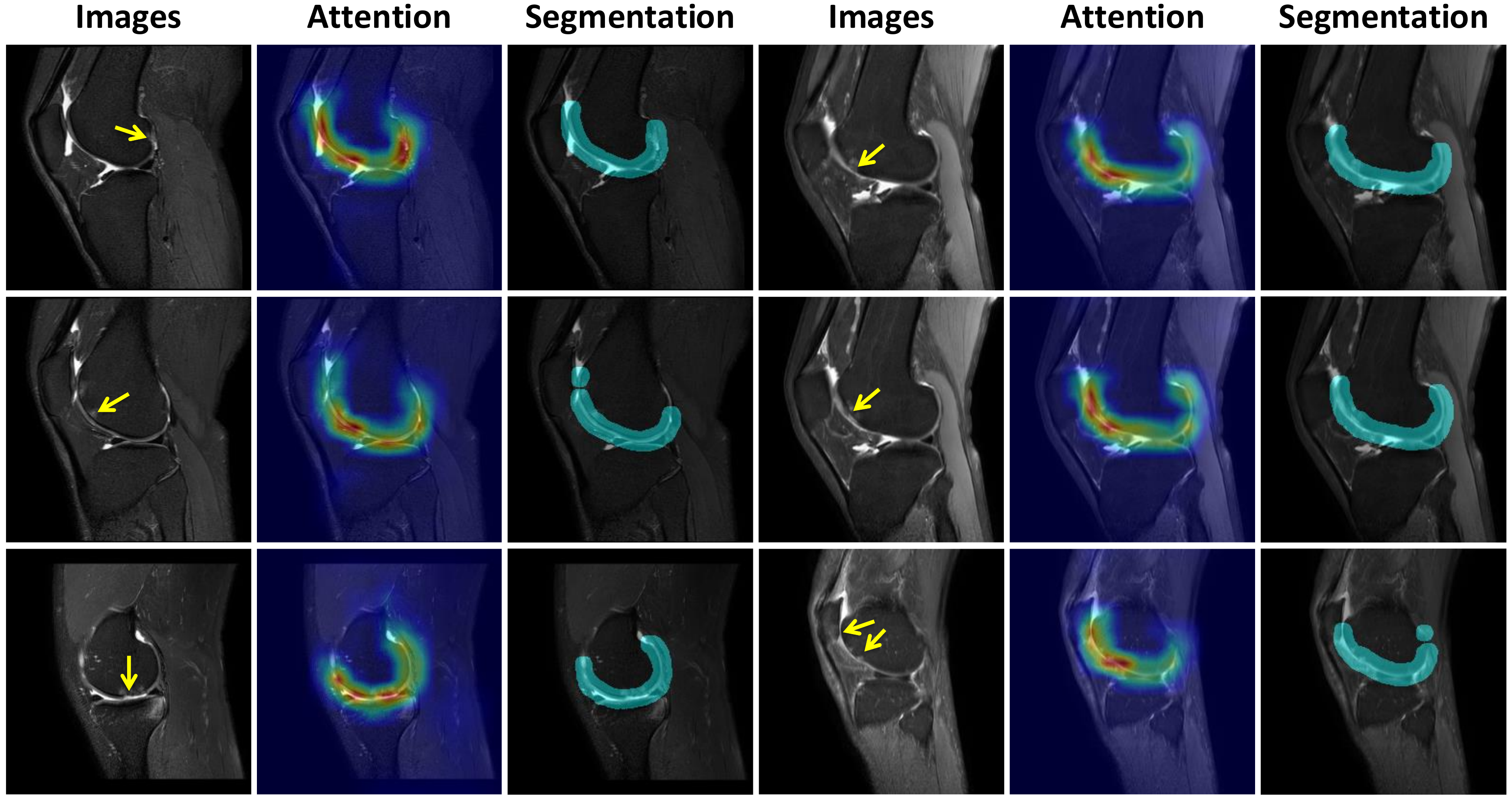}
\caption{Visualization of attention maps with the segmentation results from the knee cartilage defects diagnosis.} 
\label{fig:visualization}
\end{figure}

Fig. \ref{fig:visualization} shows three visualized results of our method when the model weight is used to make predictions on the testing set. The yellow arrows on the images indicate the specific location of knee cartilage defects, which was labeled by the experienced radiologist. It shows that the areas indicated by arrows are also highlighted by the corresponding attentions maps. More importantly, these conspicuous area in attention maps are similar to the segmentation results. Which shows that the network can classify correctly according to the accurate localization results.

\section{Conclusion}
We developed a self-enssembling semi-supervisesd network for knee cartilage defects classification and localization and proposed a dual consistency learning mechanism to coordinate the learning procedure of the student and teacher networks. Attention loss is used to not only encourage the network to yield the correct classification result, but also to provide the basis (accurate attention maps) for correct classification. Furthermore, we presented the attention consistency loss to make the general frame be consistent in the structure level. With the help of two supervised losses and dual consistency losses, our mechanism can achieve the best performance in both classification and localization tasks. The ablation experiments also confirmed the effectiveness of our method. The future works include conducting experiments in other knee datasets (\textit{e.g.}, OAI dataset) and investigating the effect of our method to other knee joint problems. 

\subsubsection{Acknowledgement}
This work was supported by the National Key Research and Development Program of China (2018YFC0116400), STCSM (19QC1400600, 17411953300, 18JC1420305), Shanghai Pujiang Program(19PJ1406800), and Interdisciplinary Program of Shanghai Jiao Tong University.

\bibliographystyle{splncs04}
\bibliography{egbib}

\end{document}